# Tunable Graphene Reflective Cells for THz Reflectarrays and Generalized Law of Reflection

Eduardo Carrasco[a)], Michele Tamagnone and Julien Perruisseau-Carrier

*Adaptive MicroNano Wave Systems, LEMA/Nanolab, Ecole Polytechnique Fédérale de Lausanne (EPFL), 1015 Lausanne, Switzerland.*

*Abstract* — A tunable graphene-based reflective cell operating at THz is proposed for use in reconfigurable-beam reflectarrays, or similarly to implement the so-called generalized law of reflection. The change in the complex conductivity of graphene when biased by an electric field allows controlling the phase of the reflected field at each element of the array. Additionally, the slow wave propagation supported by graphene drastically reduces the dimensions of the cell, which allows smaller inter-element spacing hence better array performance. An elementary cell is optimized and its scattering parameters computed, demonstrating a dynamic phase range of 300° and good loss figure for realistic chemical potential variations. Finally, a circuit model is proposed and shown to very accurately predict the element response.

Graphene is a monatomic layer of carbon atoms arranged in a honeycomb structure, which has attracted tremendous interest thanks to its unique electrical and mechanical properties[1]. From an electrical point of view, graphene is a zero-gap semiconductor, or semi-metal, whose complex conductivity nature allows the propagation of plasmonic modes at THz frequencies[2,3]. Importantly, this conductivity can be efficiently controlled via a perpendicular bias electric field. As a result graphene is envisioned for a variety of applications at THz and optical frequencies, including the possibility of dynamic tuning via the electric field effect[2-5].

In this letter, the use of graphene for realizing tunable graphene-based reflective cells, operating at THz, is proposed and investigated for the first time. Such reflective elements are needed, among other, for planar reflectarray systems, whose principle is illustrated in Fig. 1(a). A reflectarray comprises an array of reflective cells introducing a given desired phase-shift upon reflection of the wave on the surface, allowing the synthesis of a predetermined far field beam. This structure has recently attracted lots of attention at microwave frequencies, since combining the advantages of both parabolic reflectors and conventional phased arrays[6,7]. In addition to low loss, compact profile and high efficiency, reflectarrays offer the possibility to dynamically control the phase of the reflected field at the element level and therefore to reconfigure the beam.

Note that a similar concept is the so-called lens-array[8,9], where the phase-shift is applied in transmission rather than reflection. Here we will focus on demonstrating the principle in reflection, but it is also suitable for transmission. Both concepts have been extensively studied at microwave and millimetre wave bands[6-14], and more recently at optical frequencies[15], where the phenomenology was described as "generalized law of refraction and reflection", corresponding to the linear phase zone of a lens-array or a reflectarray, respectively.

[a)]Authors to whom correspondence should be addressed. Electronic mail :
eduardo.carrascoyepez@epfl.ch, julien.perruisseau-carrier@epfl.ch

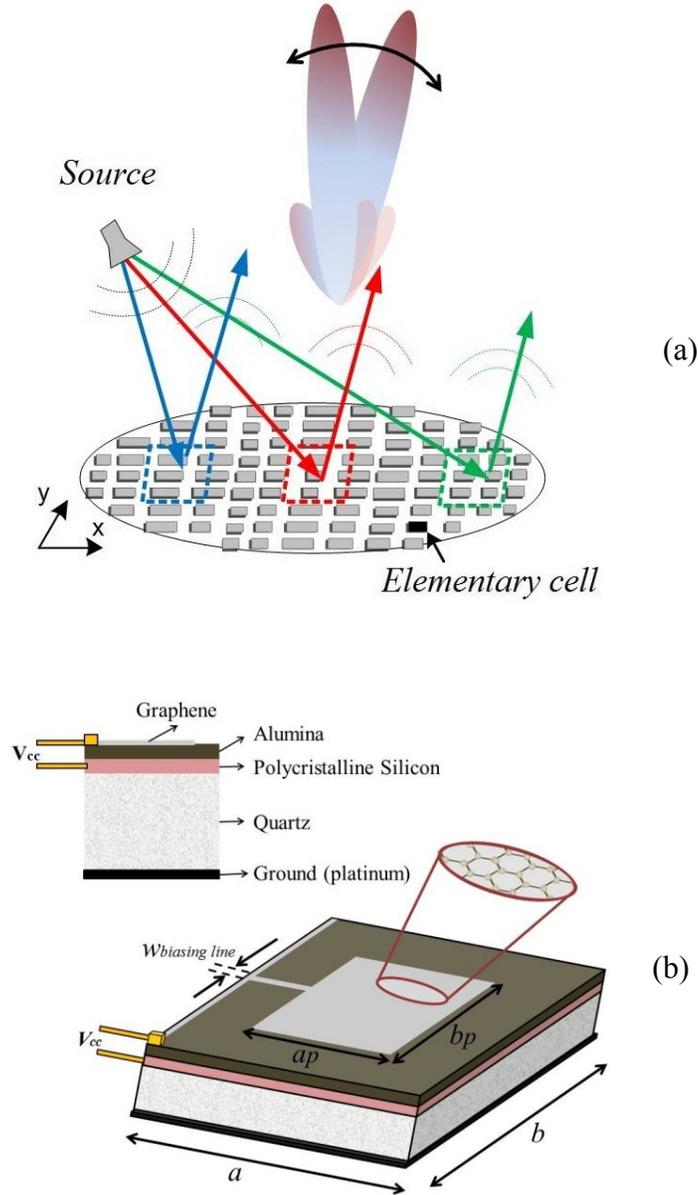

*FIG. 1: (a) General reflectarray concept. (b) Graphene reflective cell.*

Different technologies have been used to control the phase of the reflected field at micro and millimetre-waves, including semiconductor diodes[10,11] and micro-electromechanical systems (MEMS) lumped elements[12]. However, these technologies are not suited to much higher frequencies, mainly because of prohibitive loss or size. An alternative is the use of materials whose electromagnetic properties can be controlled via an external field. This is for instance the case of liquid crystals[13,14], whose anisotropic permittivity can be controlled by an applied electric field. Here, to achieve the control of reflective cells via the electric field effect on the complex conductivity of graphene is proposed, thereby benefiting from the miniaturization linked with the plasmonic propagation, easy control, and acceptable losses for THz frequencies.

Before describing the element design, the graphene conductivity model used for computations is briefly discussed. Due to its mono-atomic thickness, graphene can be accurately modeled as an infinitely thin surface of complex conductivity $\sigma$, which depends on the frequency $f$, the temperature $T$, the transport relaxation time $\tau$ and the chemical potential $\mu_c$[16]. This later parameter is affected by an externally applied electric field. In the frequency range of interest here, namely about 1-2THz, the intraband contribution of the conductivity dominates and the interband contribution can be neglected[17]. No magnetic bias is considered here. Assuming a temperature of 300K, a typical relaxation time of 1ps, and varying the chemical potential from 0eV to 0.52eV, the complex conductivity has been computed and is shown in Fig. 2. Note that in terms of surface impedance $Z_s$, the layer behaves as a constant resistance in series with an inductive reactance that increases with frequency.

A square graphene patch to be used as a reflectarray element has been designed using a grounded 30-µm $SiO_2$ substrate ($\varepsilon_r$=3.75, tan$\delta$=0.004), as shown in Fig. 1(b). In conventional reflectarray elements made of good conductors such as copper, the patch resonates when its length is in the order of a half wavelength in the effective dielectric media. Here the resonance occurs at much smaller sizes, namely, below $\lambda/10$, which is due to the well-known slow-wave propagation associated with the plasmonic mode.

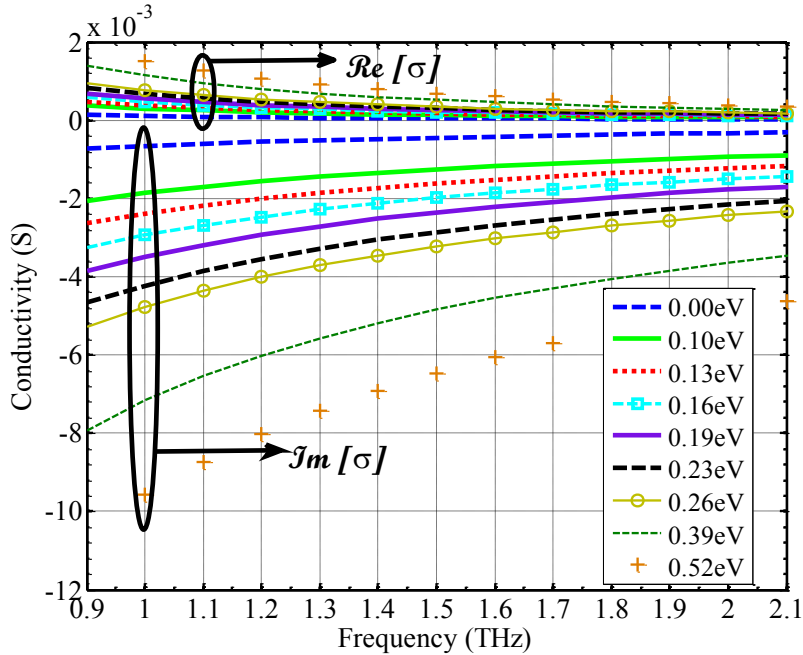

*FIG. 2  Real and imaginary part of the graphene conductivity for different values of the chemical potential ($\mu_c$).*

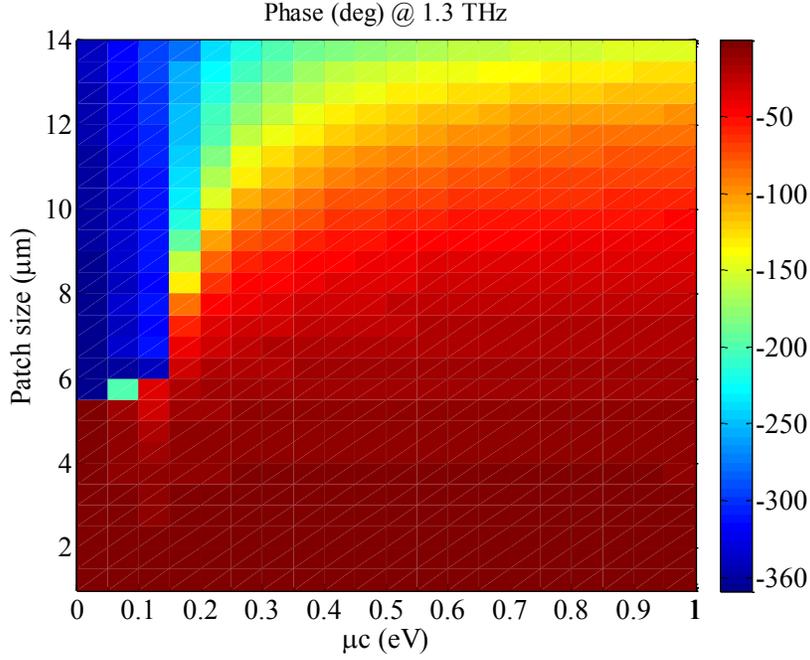

*FIG. 3   Phase shift introduced by square patches of different sizes and different values for the chemical potential ($\mu_c$), at 1.3 THz.*

The size of the unitary cell is set to $a=b=14\mu m$. The proposed electric field biasing structure is also shown in Fig. 1(b). The DC voltage is applied between a 50nm thick polycrystalline silicon layer and the graphene patch using a 10nm thick $Al_2O_3$ (Alumina, $\varepsilon_r=8.9$, $\tan\delta=0.01$) as spacer. The graphene patch is connected through a thin line also made of graphene (width=50nm) to an electrode located at the border of the overall structure. The thin polycrystalline and $Al_2O_3$ layers, as well as the thin biasing lines have negligible impact on the element response. In this way all elements in the same row are interconnected and will be subject to the same bias voltage, while the different rows can be independently controlled, allowing scanning the radiation maxima in a determined plane. More complex biasing schemes can of course be implemented for more flexibility, for instance controlling each cell individually.

The phase of the reflected wave has been calculated for different sizes of the patch for normal incidence, varying at the same time the chemical potential $\mu_c$. The computations were carried out using a full-wave tool (CST Microwave Studio®), taking into account the inter-element mutual coupling by a local periodicity assumption, as usually in reflectarrays. Fig. 3 shows the phase variation as a function of the two parameters at 1.3THz. The phase variation for $\mu_c$ greater than 0.52eV is practically negligible for all the patch sizes, thus it can be considered as the maximum useful value of chemical potential here. Importantly this maximum is significantly below the chemical potential observed in different works[18] and is thus achievable in practice. On the other hand, sufficient phase variation must be available for covering most of the ideal 360° phase range. In practice, a phase range superior to 270° allows providing good performance[6] and here the maximum phase variation is obtained for patches of 10μm, yielding a range of 300°.

Fig. 4 shows the phase and amplitude of the reflection coefficient for a normally-incident plane wave impinging on the proposed element ($a=b=14\mu m$, $a_P=b_P=10\mu m$), from 0.9 THz to 2.1THz. Because of the symmetry of the element, an identical response is obtained for the orthogonal polarization. A variation in the chemical potential allows tuning the phase in a range of 300° in the whole band, with very linear phase variation in the band from 1.2THz to 1.5THz (namely, large bandwidth[6]). The loss of the element varies between 0.5B and 6dB on the whole range between 1.1THz and 1.6THz, which is another very promising performance at such frequencies.

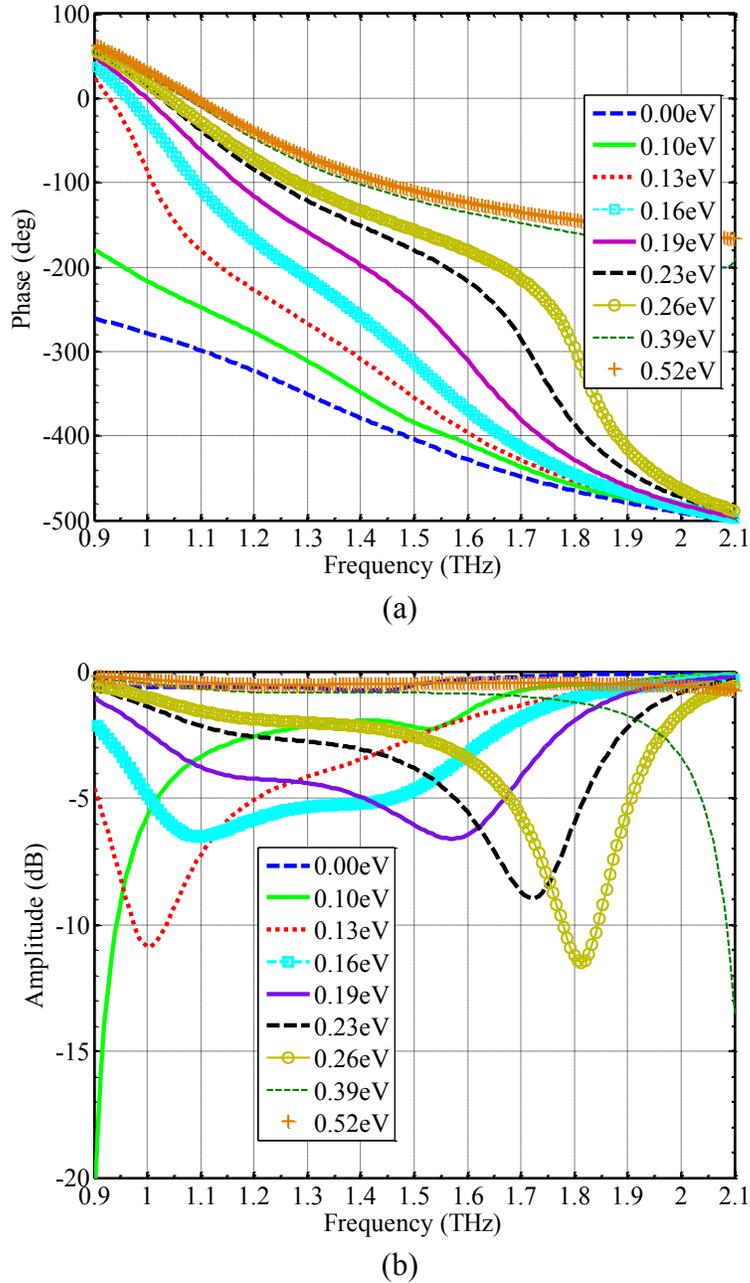

*FIG. 4   Reflection coefficient in free-space, as a function of the frequency, for a 10μm-side square patch made of graphene varying the chemical potential ($\mu_c$). (a) Phase. (b) Amplitude.*

The proposed reflecting cell can be modelled by a physical equivalent circuit such as shown in Fig. 5, where the graphene patch between two stratified media (air-quartz) is represented as an *R-L-C* circuit in parallel with the grounded substrate and referred to the intrinsic impedance of air $\eta_0$. The values of *R*, *L* and *C* have been extracted from the reflection and transmission scattering parameters of the graphene patch at the boundary between air and substrate. These parameters depend on the size of the patch and the chemical potential, and Table I shows the extracted *R*, *L* and *C* values for various representative cases. First, it is observed that both resistance and inductance reduce as the chemical potential $\mu_c$ increases. This trend is perfectly explained by the evolution of the real and imaginary part of the graphene conductivity when the chemical potential is varied (see Fig. 2). Second, the capacitive component of the equivalent impedance is almost independent of the chemical potential, but increases significantly with the patch size. Both trends are again physically intuitive, since the capacitive component is mostly due to the electrostatic capacitive coupling between conductive patches.

TABLE I
VALUES OF THE CIRCUIT ELEMENTS FOR A GRAPHENE PATCH

| PATCH SIDE (µM) | ELECTRIC PARAMETER | GRAPHENE | | |
|---|---|---|---|---|
| | | 0.00 eV | 0.19 eV | 0.52 eV |
| 3.5 | R (Ω): | 8854 | 799 | 287.5 |
| | L (pH): | 3495 | 803 | 294 |
| | C (fF): | 0.006 | 0.004 | 0.005 |
| 7.0 | R (Ω): | 1861 | 214 | 72.5 |
| | L (pH): | 862.1 | 208.9 | 75.07 |
| | C (fF): | 0.068 | 0.042 | 0.042 |
| 10.0 | R (Ω): | 861 | 122.5 | 35.4 |
| | L (pH): | 435 | 99.69 | 36.22 |
| | C (fF): | 0.148 | 0.137 | 0.133 |
| 14.0 | R (Ω): | 236 | 44.52 | 16.27 |
| | L (pH): | 235 | 44.44 | 16.24 |
| | C (fF): | VERY HIGH | VERY HIGH | VERY HIGH |

Finally Fig. 6 compares the reflection coefficients obtained by the circuit model with full-wave simulations. The results are shown for different patch sizes, in the case $\mu_c = 0.19$eV. An excellent agreement is observed in both phase and amplitude.

In summary, a reflectarray element based on a square graphene patch is proposed, demonstrating the possibility of implementing phase control at THz through chemical potential variations. Very promising performance was obtained in terms of phase range, loss, and bandwidth, paving the way for future theoretical and experimental developments in the use of graphene for free-space wave propagation control.

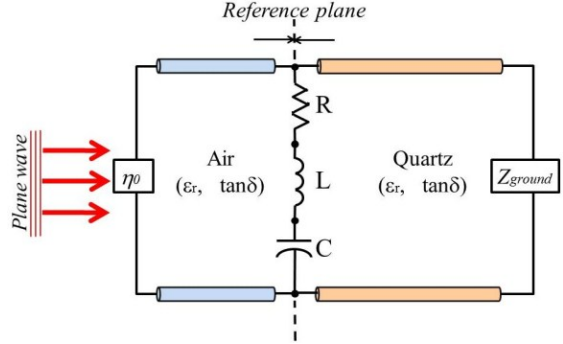

FIG. 5 Equivalent circuit for the proposed element.

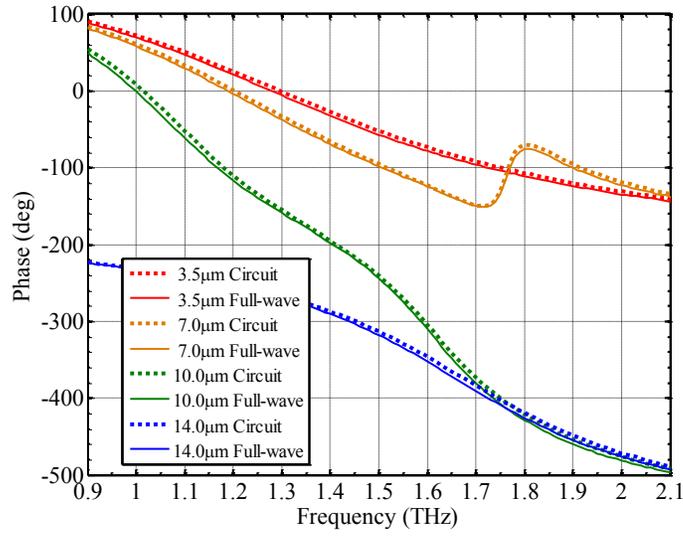

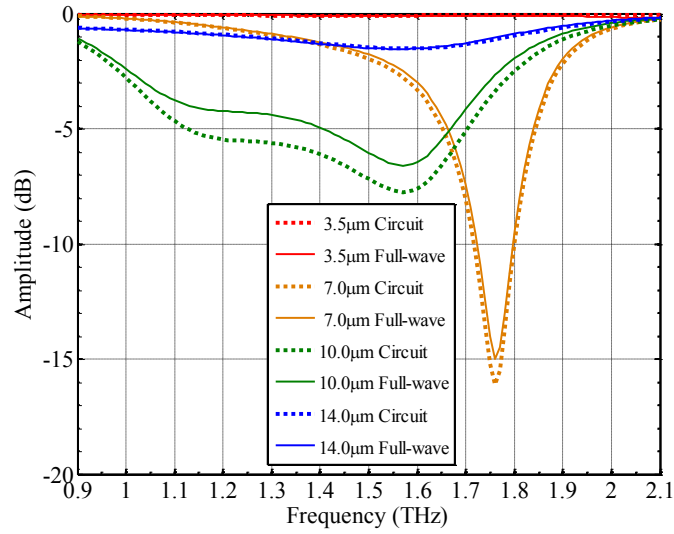

FIG. 6 Reflection coefficient of the graphene element, comparison between equivalent circuit and full-wave simulation, for $\mu_c$=0.19 eV. (a) Phase. (b) Amplitude.

This work was partially supported by European Union (FP7) under grant 300934 (IEF Marie-Curie Project *RASTREO*), by the Swiss National Science Foundation (SNSF) under grant no. 133583, and by the Hasler Foundation, Project 11149.